\PassOptionsToPackage{numbers,sort&compress}{natbib}

\documentclass{article}

 \usepackage[preprint]{neurips_2025}

\usepackage[utf8]{inputenc} % allow utf-8 input
\usepackage[T1]{fontenc}    % use 8-bit T1 fonts
\usepackage{hyperref}       % hyperlinks
\usepackage{url}            % simple URL typesetting
\usepackage{booktabs}       % professional-quality tables
\usepackage{amsfonts}       % blackboard math symbols
\usepackage{nicefrac}       % compact symbols for 1/2, etc.
\usepackage{microtype}      % microtypography
\usepackage{xcolor}         % colors
\usepackage{amsmath} % Required for the 'cases' environment and \text{} in math mode
\usepackage{xspace}
\usepackage{listings}
\usepackage{graphicx}
\usepackage{enumitem}  
\usepackage[most]{tcolorbox}
\usepackage{ragged2e} 
\usepackage{subcaption}

\usepackage{hyperref} % hyperref should usually be loaded before setcitestyle for colors
\setcitestyle{numbers,square,citecolor=green}

\hypersetup{
    colorlinks=true,  
    linkcolor=red,      
    citecolor=green,    
    filecolor=magenta,  
    urlcolor=magenta    
}

\renewcommand{\cite}{\citep}

\setlist[itemize]{leftmargin=1.5em} 
\setlist[enumerate]{leftmargin=1.5em} 

\newtcolorbox{codebox}{
  colback=black!5!white,
  colframe=black!75!black, 
  fontupper=\ttfamily, 
  boxrule=0.5pt, 
  arc=3pt,
  boxsep=5pt, 
  left=5pt,
  right=5pt, 
  top=5pt, 
  bottom=5pt, 
  before upper={\RaggedRight}, 
}

\newcommand{\sysmem}{ActTree\xspace}
\newcommand{\sysbench}{MobiFlow\xspace}
\newcommand{\FEH}[1]{\textcolor{black}{#1}}
\title{MobiAgent: A Systematic Framework for Customizable Mobile Agents}

\author{Cheng Zhang, Erhu Feng$^*$, Xi Zhao, Yisheng Zhao, Wangbo Gong,\\ 
        \textbf{Jiahui Sun}, \textbf{Dong Du}, \textbf{Zhichao Hua}, \textbf{Yubin Xia}, \textbf{Haibo Chen}\\
	\emph{Institute of Parallel and Distributed Systems (IPADS)}, \\
	\emph{Shanghai Jiao Tong University}
} % removed for anonymity

\begin{document}

\maketitle

\begin{abstract}
  With the rapid advancement of Vision-Language Models (VLMs), GUI-based mobile agents have emerged as a key development direction for intelligent mobile systems. However, existing agent models continue to face significant challenges in real-world task execution, particularly in terms of accuracy and efficiency. To address these limitations, we propose \textbf{MobiAgent}, a comprehensive mobile agent system comprising three core components: the MobiMind-series agent models, the AgentRR acceleration framework, and the \sysbench benchmarking suite. Furthermore, recognizing that the capabilities of current mobile agents are still limited by the availability of high-quality data, we have developed an AI-assisted agile data collection pipeline that significantly reduces the cost of manual annotation. Compared to both general-purpose LLMs and specialized GUI agent models, MobiAgent achieves state-of-the-art performance in real-world mobile scenarios.
\end{abstract}

\section{Introduction}

With the rise of Vision-Language Models (VLMs) and the growing adoption of mobile agents, we have observed that agents leveraging GUI/XML understanding have immense application potential on real mobile devices. 
Unlike traditional mobile intelligent assistants, which primarily rely on rule-based and intent-driven API calls, GUI/XML-based agents are capable of adapting to any mobile application without requiring additional adaptation from app developers. As a result, an increasing number of vendors have introduced their own agent models for mobile devices, including specialized agent models such as UI-TARS~\cite{qin2025uitarspioneeringautomatedgui}, MobileAgent~\cite{wang2024mobileagentv2mobiledeviceoperation}, CogAgent~\cite{hong2024cogagentvisuallanguagemodel}, and etc.~\cite{lin2024showuivisionlanguageactionmodelgui,zhang2023appagentmultimodalagentssmartphone,wu2024osatlasfoundationactionmodel,zhang2024lookscreensmultimodalchainofaction,gou2024uground,10.1145/3636534.3690682,10.1145/3636534.3649379,cheng2024seeclickharnessingguigrounding,zheng2024seeact}, as well as more general-purpose models like GPT-5~\cite{GPT-5} and Gemini 2.5-pro~\cite{gemini-2.5-pro}.

Despite this progress, significant challenges remain when deploying these agent models in real-world environments. For example, existing agents often exhibit low task completion rates, slow response times, and limited ability to handle unexpected situations.

We first introduce the MobiMind-series models, which employ a multi-role architecture consisting of three distinct components: Planner, Decider, and Grounder. This design decouples task planning, reasoning, and execution, thereby facilitating seamless integration with various backend operation modes (e.g., GUI, XML, etc.).
In addition, we propose an agent acceleration framework: AgentRR~\cite{feng2025experiencepracticellmagents}. AgentRR records execution traces from the agent’s operation and abstracts them into multi-level experiences. A lightweight latent memory model determines whether the agent can leverage past experiences, thereby reducing the computational burden on VLMs/LLMs during task execution. With the AgentRR framework, the mobile agent can continuously enhance its efficiency and accuracy on recurring tasks.

To accurately evaluate the performance of our mobile agent models in real-world scenarios, we have surveyed existing mobile benchmark frameworks, including AITW~\cite{10.5555/3666122.3668731}, ANDROIDCONTROL~\cite{li2024effectsdatascaleui}, AndroidArena~\cite{xing2024understandingweaknesslargelanguage}, AndroidWorld~\cite{rawles2025androidworlddynamicbenchmarkingenvironment}, GUIOdyssey~\cite{lu2025guiodysseycomprehensivedatasetcrossapp}, A3~\cite{chai2025a3androidagentarena}, and etc. \cite{li2025screenspotproguigroundingprofessional,zhang2024mobileenvbuildingqualifiedevaluation,kapoor2024omniactdatasetbenchmarkenabling,lee2024benchmarking,pan2024autonomous}. However, in practical usage scenarios, the presence of variable environmental factors, multiple correct execution trajectories, and the lack of deterministic verification mechanisms make it challenging for current benchmark frameworks to provide accurate assessments.
To address these challenges, we design a novel mobile benchmarking framework: \sysbench, based on Directed Acyclic Graphs (DAGs) and milestone events. \sysbench defines multiple correct trajectories and employs a multi-level verification mechanism (e.g., text/icon matching, OCR, and LLM-based analysis) to enable precise and fine-grained evaluation of agent tasks. Furthermore, to mitigate the impact of environmental variability, \sysbench enables replay-based evaluation by allowing different agents to execute pre-recorded static traces. This approach ensures a more controlled and deterministic assessment of agent performance.

Our experimental results demonstrate that, on real-world mobile scenarios (as evaluated by the \sysbench Benchmark), the combination of MobiMind-Decider-7B and MobiMind-Grounder-3B outperforms general-purpose large language models (e.g., Gemini-2.5 Pro, GPT-5) as well as other specialized mobile agent models (such as UI-TARS-1.5-7B). Furthermore, MobiMind exhibits superior instruction-following capabilities, generates higher-quality reasoning, and achieves more reliable task termination.
\section{Real-world Trajectory Collection}
\label{sec:data}

In this section, we present a framework for collecting real-world GUI task trajectories with rich contextual information and high data quality, while requiring minimal human efforts.
The collected trajectories are further refined and used to train our agent models, enabling them to learn from diverse and realistic user interactions.

\subsection{Agent Trajectory Collection}
\label{sec:collection}

To ensure that the collected trajectories accurately reflect real-world user behavior, we developed a lightweight action recording tool for smartphones. During data collection, annotators interact with this tool, which captures and logs every user action before forwarding it to the device. The tool renders all bounding boxes (bbox) of interactive elements on the screen based on XML files that describe UI hierarchies. If a bbox is missing due to incomplete XML data, we employ OmniParser~\cite{lu2024omniparserpurevisionbased} to regenerate the corresponding bbox.

The recorded action space includes:

\begin{itemize}
    \item \texttt{Click(bbox:~List[int])}: Click at the center point of \texttt{bbox} using absolute coordinates.
    \item \texttt{Input(text:~str)}: Input \texttt{text} to a currently activated input area.
    \item \texttt{Swipe(direction:~Literal["UP","DOWN","LEFT","RIGHT"])}: Swipe to \texttt{direction}.
    \item \texttt{Done()}: Signify the completion of the task.
\end{itemize}

For rare but important cases that are difficult to capture with the recording tool efficiently, such as closing a pop-up ad or waiting for it to disappear, we create a separate single-step dataset containing these cases and add an new \texttt{wait} action to the action space:

\begin{itemize}
    \item \texttt{Wait(sec:~int)}: Wait for \texttt{sec} seconds before the next action.
\end{itemize}

For simpler tasks which a powerful pretrained VLM (e.g. Gemini-2.5) can handle, we directly prompt the VLM to execute these tasks and record the trajectories with a uniform format, which further improves the efficiency of data collection. 

\subsection{Agent Reasoning Reconstruction}

The recorded trajectories only contain low-level information (e.g., raw coordinates or bounding boxes), 
lacking high-level semantics (e.g., task planning and environment observation) which is crucial for training intelligent and robust agent models.
To address this issue, we propose a VLM-based reasoning reconstruction approach which leverages Gemini-2.5
to reconstruct the agent's reasoning process based on original trajectories. 

Specifically, we prompt the VLM to examine each action in the trajectory and provide a detailed reasoning, simulating an agent who executes the task following the ReAct~\cite{yao2023react} paradigm.
Furthermore, for each concrete operation (e.g., click bbox), we also utilize the VLM to provide an action primitive (e.g., click search button) that abstractly describes this operation.

\begin{figure}[t]
    \centering
    \includegraphics[width=\linewidth]{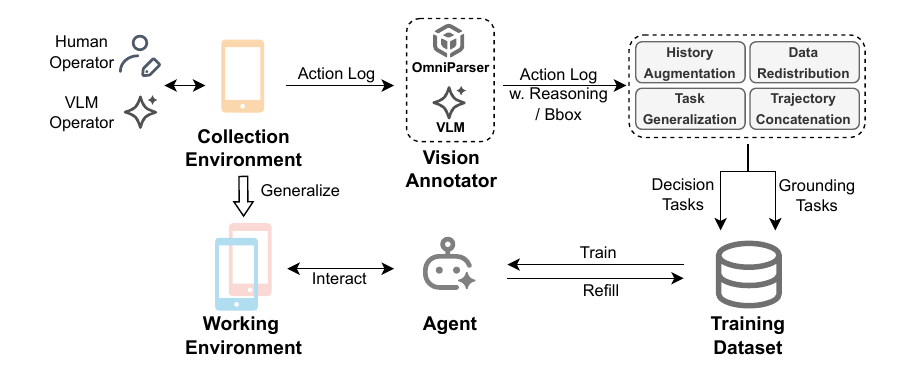}
    \caption{\textbf{Data Collection Pipeline with Agent Self-evolving}}
    \label{fig:data_pipeline}
\end{figure}

\subsection{Post-collection Data Refinement}

Initially collected trajectories may contain errors, have unbalanced distributions, or lack diversity. 
Besides applying a series of rule-based automated checks and manual reviews to remove erroneous or inconsistent samples, we further adopt the following data refinement strategies:

\begin{itemize}
    \item \textbf{Task Concatenation:} We concatenate temporally dependent trajectories to form new trajectories corresponding to more complex tasks. In this way, the dataset diversity is boosted without extra efforts.
    \item \textbf{Data Redistribution:} Different trajectories tends to share similar prefixes while having diverse suffixes. When constructing training data, we perform sampling on the prefix actions while keeping all the suffix actions in the dataset, ensuring a more balanced distribution.
    \item \textbf{History Augmentation:} Besides samples with complete historical information, we also include samples with only partial history in training data. This approach avoids the model from overfitting to specific history patterns.
    \item \textbf{Prompt Generalization:} In real-world scenarios, agent's input task descriptions varies across different users. However, each trajectory obtained during the initial data collection has only one corresponding task description, which can lead to weaker generalization abilities of the agent. Therefore, we assign more semantically equivalent task desciptions to every trajectory to build a more comprehensive training dataset.
    \item \textbf{Corner-case Enhancement:} With the single-step dataset obtained in Secion \ref{sec:collection}, we enhance the agent's ability to handle more complicated corner cases (e.g., closing pop-up ads, terminating current task or returning to the previous page upon execution errors) by mixing single step desicion with history-based decision at training time. Experiments show that at test time, decider has stable capability of handling these problems at any time step without being interfered by the action history.
\end{itemize}
\section{Training}
\label{sec:train}

We post-train decider and grounder on the basis of Qwen2.5-VL-7B and Qwen2.5-VL-3B~\cite{bai2025qwen25vltechnicalreport}, respectively. 
% The post-training process incorporates 3 phases: a self-evolving SFT phase, an error-correcting DPO phase, and a two-stage GRPO phase.
The post-training process incorporates 2 phases: a warm-up SFT phase and a two-stage curriculum GRPO~\cite{shao2024deepseekmathpushinglimitsmathematical} phase.

\subsection{Warm-up SFT}

In SFT phase, we use a subset of the training data with the following prompt templates to train decider and grounder:

\begin{codebox}
Decider:\\
<image>\\
You are a phone-use AI agent.~Now your task is "\{task\}".\\
Your action history is:\\
\{history\}\\
Please provide the next action based on the screenshot and your action \\
history.~You should do careful reasoning before providing the action.\\
Your action space includes:\\
- Name:~click, Parameters:~target\_element(a high-level description of the UI element to click).\\
- Name:~swipe, Parameters:~direction(one of UP, DOWN, LEFT, RIGHT).\\
- Name:~input, Parameters:~text(the text to input).\\
- Name:~wait, Parameters:~(no parameters, will wait for 1 second).\\
- Name:~done, Parameters:~(no parameters).\\
Your output should be a JSON object with the following format:\\
\{"reasoning":~"Your reasoning here",~"action":~"The next action (one of click, input, swipe, wait, done)", "parameters":~\{"param":~"value"\}\}
\end{codebox}

\begin{codebox}
Grounder:\\
<image>\\
Based on the screenshot, user's intent and the description of the target UI element, provide the element's bounding box using absolute coordinates.\\
User's intent:~\{reasoning\}\\
Target element's description:~\{target\_element\}\\
Your output should be a JSON object with the following format:\\
\{"bbox":~[x1, y1, x2, y2]\}
\end{codebox}

This warm-up phase empowers both models with basic format following and GUI agent capabilities, avoiding the reward signals in the subsequent RL phase being overly sparse which causes harm to training effectiveness.

\subsection{Two-stage Curriculum GRPO with Self-evolution}

Since MobiAgent adopts a multi-agent architecture, we simplify the multi-agent reinforcement learning process by incorporating 2 separate stages in GRPO phase to train the grounder and decider independently.

\textbf{Grounding GRPO. } Firstly, we train the grounder with a rule-based reward function using the UI grounding dataset in Section \ref{sec:data}. The reward function consists of two parts: an IoU reward $R_{\mathrm{iou}}$ and a center-point reward $R_{\mathrm{center}}$, reflecting the precision of the bounding box prediction of the grounder: 

\begin{gather*}
    R_{\mathrm{iou}} = 
    \begin{cases}
    \alpha, & \mathrm{IoU}\left(b_\mathrm{pred},b_{gt}\right) > \beta, \\
    0, & \text{otherwise}.
    \end{cases}
    ,\;
    R_{\mathrm{center}} = 
    \begin{cases}
    1-\alpha, & \mathrm{Center}\left(b_\mathrm{pred}\right) \text{\;is inside\;} b_{gt}, \\
    0, & \text{otherwise}.
    \end{cases},
    \\
    R=R_{\mathrm{iou}}+R_{\mathrm{center}},
\end{gather*}

where $b_\mathrm{pred},b_{gt}$ are the predicted and ground-truth bounding boxes, $\mathrm{IoU}\left(\cdot\right)$ is the Intersection over Union function, and $\alpha,\beta$ are hyperparameters.

\textbf{Grounder-as-RM GRPO. } For click actions, since the decider's output only contains a high-level description of the target element, it is unfeasible to evaluate the its quality by simply comparing it to the ground-truth. However, the grounder is able to map the high-level semantics to low-level and judgable actions. Therefore, after a high-precision grounder is obtained in the previous stage, we then let it serve as an outcome reward model (ORM) for training the decider. 

Specifically, for each response generated in the rollout stage, if the action is a non-click operation, we apply a binary reward $R_{\mathrm{content}}$ by exact matching:

$$
R=R_{\mathrm{content}} = 
\begin{cases}
1, & a_\mathrm{pred}=a_{gt}, \\
0, & \text{otherwise}.
\end{cases},
$$

where $a_\mathrm{pred},a_\mathrm{gt}$ are the predicted and ground-truth actions. 
For click action, the grounder predicts a bounding box of the target element in decider's response, and directly returns $R=R_{\mathrm{center}}$ as the final reward.

\textbf{Curriculum Learning. }In Grounder-as-RM stage, we further adopt a curriculum learning~\cite{10.1145/1553374.1553380}-based training approach to train the decider: Each action in the training dataset is labeled a binary difficulty with a series of predefined rules (e.g., action type frequency or manually assigned importance). In early training, each batch in model rollout only contains ``easy'' samples, and we gradually mix more ``difficult'' samples in the batches as the training progresses. This approach further addresses sparse rewards, accelerates model convergence and improves training robustness.

\textbf{Self-evolution. }To avoid unexpected agent behavior at test time due to error accumulation in the agents' input context (i.e., the \texttt{history} part in decider's input and the \texttt{reasoning} part in grounder's input), 
we align the distribution of agent input at training time and test time via \textbf{training-time-test}. 
Specifically, after each round of training, we test the two models with new tasks and collect the test-time traces. Then we examine and correct the unsuccessful traces, and merge the corrected traces with successful traces to become a part of the training data of the next round, enabling the agent to self-evolve in an iterative way.
\section{AgentRR: Record-Replay-based Agent Acceleration with Multi-level Experiences}

In this section, we introduce a general-purpose agent acceleration framework: \textbf{AgentRR}, as well as our specific implementation in mobile scenario. AgentRR seamlessly supports any mobile agent models and applications without the need for additional development for specific tasks. Empirical evaluations show that \sysmem enables GUI agents to achieve execution efficiency comparable to that of API agents, while preserving the generalization capability of GUI agents.

\begin{figure}[t]
    \centering
    \includegraphics[width=\linewidth]{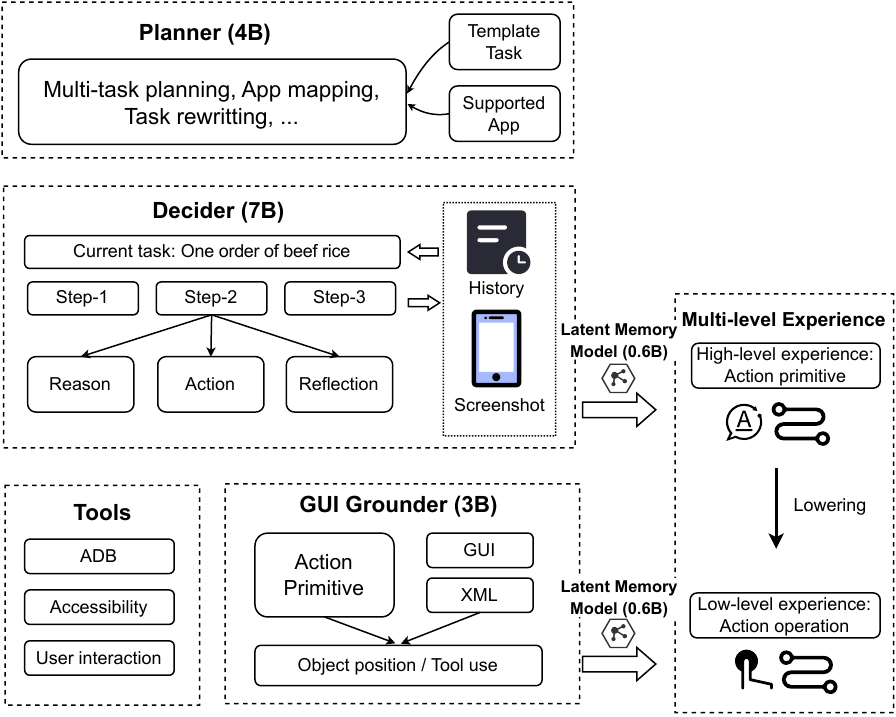}
    \caption{\textbf{Multi-Agent Architecture using the AgentRR Framework}}
    \label{fig:agentRR}
\end{figure}

\subsection{Multi-level Experiences}
As shown in Figure~\ref{fig:agentRR}, MobiAgent adopts a multi-agent architectural design, including a Planner (4B), a Decider (7B), and an Grounder (3B).  
The execution of a complete task requires the collaborative operation of these multiple models/agents.  
We record the output of each model as an experience, such that the outputs from different models form a multi-level experience hierarchy.  
The Grounder outputs concrete operation actions, such as the bounding box of the clicked object, representing the lowest-level experiences.  
The Decider outputs action primitives described in natural language.  
The Planner outputs the task plan, corresponding to high-level experiences.  
Lower-level experiences can be executed with higher efficiency, but typically offer weaker generalization, while higher-level experiences exhibit stronger generalization capabilities but are less efficient to execute (as they still require model inference).

To address this challenge, we propose a record-and-replay system specifically designed for agent frameworks.
The core idea behind it is to efficiently determine which actions can be directly reused, without compromising the agent’s generalization capabilities. This mechanism is analogous to human latent memory, enabling swift execution of familiar actions while preserving the flexibility required for new situations.

\subsection{Prefix Reusablity of GUI tasks}

We observe the prefix reusablity of GUI tasks, which relies on the following insights:

\begin{itemize}
    \item \textbf{I1:} The execution trajectories of similar tasks share identical prefixes. 
    \item \textbf{I2:} Since the agent's planning can be directly derived from task description, 
    \FEH{the length of shared trajectories (i.e., prefix actions) between two tasks can be predicted based on the similarity of their task descriptions.}
\end{itemize}

According to \textbf{I1}, task trajectories can be cached using a tree structure, where similar tasks share the same path from the root node to their common ancestor node.
According to \textbf{I2}, \FEH{the agent can determine whether to reuse actions from historical tasks at the current step through simple and efficient semantic matching, 
thereby avoiding redundant model inference.}
By exploiting prefix reusability, we can efficiently records execution trajectories and attempts to replay cached actions for every execution step, thereby reducing the end-to-end task execution latency.

\subsection{\sysmem Structure}

\begin{figure}[t]
    \centering
    \includegraphics[width=\linewidth]{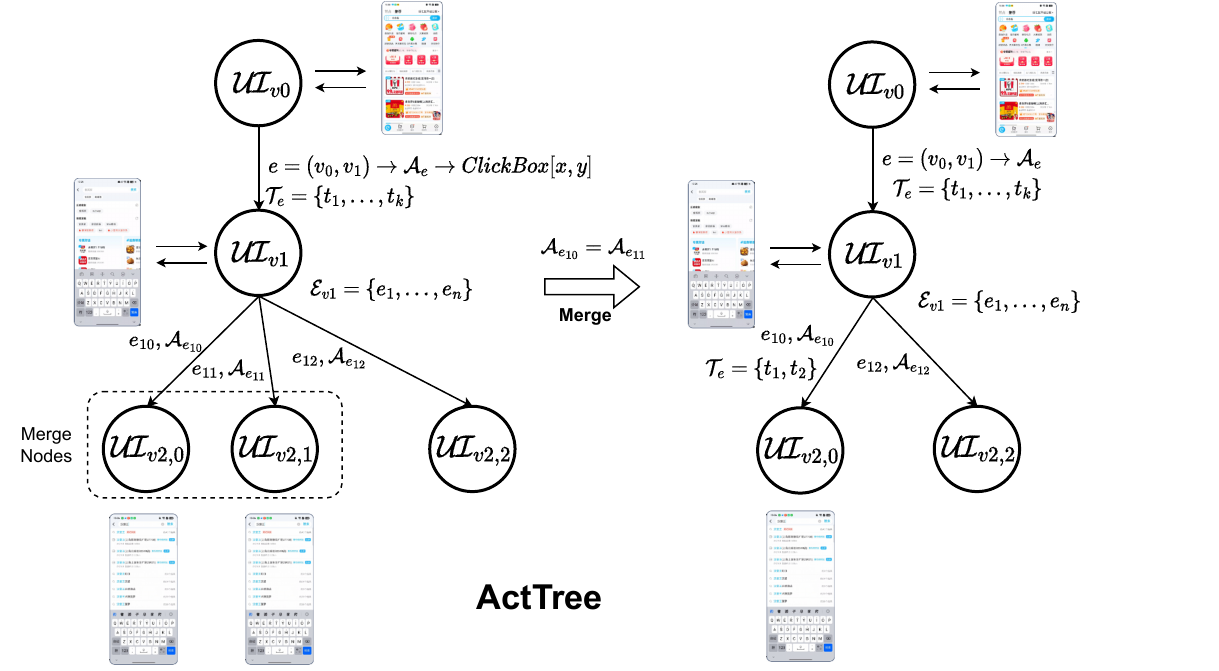}
    \caption{\textbf{Construction of the \sysmem Structure During Mobile Task Execution}}
    \label{fig:acttree}
\end{figure}

\FEH{\sysmem is a concrete implementation mechanism that accelerates mobile tasks by leveraging the principles of AgentRR. As shown in Figure~\ref{fig:acttree} \sysmem  maintains a tree-structured representation analogous to UI Transition Graphs. Each node $v \in \mathcal{V}$ in the tree corresponds to a unique UI state, denoted as $\mathcal{UI}_u$, and maintains a set of outgoing edges $\mathcal{E}_u = \{e_1, ..., e_n\}$. Each edge $e = (u, v) \in \mathcal{E}$ represents a state transition, which is characterized by: (1) an action $\mathcal{A}_e$ that triggers this transition; and (2) a task list $\mathcal{T}_e = \{t_1, ..., t_k\}$, which records all historical tasks that executed action $\mathcal{A}_e$ to transition from $\mathcal{UI}_u$ to $\mathcal{UI}_v$.}

\FEH{The construction of the ActTree proceeds incrementally during the task executions.  
When a new action is performed, we first create a new node $v$ to represent the resulting UI state: $\mathcal{UI}_v$, along with a corresponding edge $e_{\mathrm{new}} = (u, v)$ annotated with the executed action $\mathcal{A}_{e_{\mathrm{new}}}$ and the current task description $t_{\mathrm{new}}$.  
To maintain the compactness of the ActTree, the system compacts the set of outgoing edges $\mathcal{E}_u = \{e_1, \ldots, e_n\}$ from node $u$.  
For example, if there exists an edge $e_{\mathrm{old}} \in \mathcal{E}_u$ whose action $\mathcal{A}_{e_{\mathrm{old}}}$ matches the new action $\mathcal{A}_{e_{\mathrm{new}}}$, the system merges these two edges.  
The merge operation combines the task lists of the two edges by updating $\mathcal{T}_{e_{\mathrm{merge}}} \leftarrow \mathcal{T}_{e_{\mathrm{old}}} \cup \{t_{\mathrm{new}}\}$, while leaving the remaining edge structure unchanged.}

\FEH{To support more complex agent scenarios, such as recursive task invocation and collaboration among multiple applications, 
we further extend the ActTree structure to allow certain backtracking edges.
For example, edges from child nodes back to parent nodes enable the agent to return to the parent's UI state when an erroneous action is executed
or when recursive tasks are invoked.
Additionally, we also introduce cross-subtree edges with the additional transferred message to facilitate coordination across multiple tasks.}

In addition to \sysmem itself, we incorporate a \textbf{Tracer} component that is bound to \sysmem to monitor task execution.  
During the execution of a new task, Tracer advances to a new node along with each UI transition.  
At each step, if Tracer identifies an existing outgoing edge $e$ whose action $\mathcal{A}_e$ can be reused for the current task (i.e., a cache hit), it replays $\mathcal{A}_e$ without invoking the agent model.  
Otherwise, the Tracer invokes the agent model to generate the next action.

\subsection{Latent Memory Models}
\label{sec:cache_hit}

\FEH{When using AgentRR framework to accelerate agent execution, a key challenge is determining how many prefix actions can be reused. To address this, we propose the ``Latent Memory'' model, which efficiently computes the similarity between two tasks and thereby determines the extent to which prefix actions can be reused.}
\FEH{Specifically, the "Latent Memory" model is implemented in the form of the task embedding and task reranking models. These models typically have significantly fewer parameters than the planner or decider, resulting in lower response latency.}

\subsubsection{Task Embedding}

\FEH{The task embedding model is responsible for computing the similarity between two tasks, which is measured as the cosine similarity between their respective embedding vectors.}
\FEH{Below, we provide the formal representation of the embedding at a specific layer of the \sysmem for a given task.}

\begin{equation*}
    v_{i}^{(l)} = f(T_i,l) \in \mathbb{R}^d.
\end{equation*}

\noindent\textbf{Where:}
\begin{itemize}
    \item $T_i$ represents the \textbf{$i$-th task}.
    \item $f$ is an \textbf{instruction-aware task embedding model}, and $f(T_i,l)$ maps $T_i$ to a vector for the \textbf{$l$-th layer} of the \sysmem. $l$ is in the instruction part of $f$'s input.
    \item $v_{i}^{(l)}$ is the \textbf{embedding vector} corresponding to the task $T_i$ and the layer $l$.
    \item $d$ is the \textbf{dimension} of the embedding vector, and $\mathbb{R}^d$ represents the $d$-dimensional real-valued vector space where the embedding vector resides.
\end{itemize}

\FEH{To determine whether the action prefix of task $T_i$ can be reused by task $T_j$, we compare the cosine similarity of task embedding $S(v_{i}^{(l)}, v_{j}^{(l)})$ with a predefined similarity threshold $\tau_1$.}
%Since action reuse needs to be evaluated at each layer of the ActTree, this threshold is defined on a per-layer basis.
\FEH{The stage-1 decision of whether $T_i$ can reuse the actions of $T_j$ at the $l$-th layer in \sysmem is formally expressed as:}

\[
\text{Embed-Reuse}^{(l)}(T_i, T_j) =
\begin{cases}
  \text{True}, & \text{if } S(v_{i}^{(l)}, v_{j}^{(l)}) \ge \tau_1 \\
  \text{False}, & \text{otherwise}
\end{cases}.
\]

\subsubsection{Task Reranking}
\FEH{Another approach to implementing the latent model is to leverage task reranking. Given two candidate tasks, the latent model determines whether the action at a specific layer of \sysmem can be reused, and directly return a score (the logit of "yes" token) between $0$ and $1$. The task reranking mechanism serves as a complementary method to task embedding. In our implementation, we first utilize the task embedding model to compute similarity scores between the current task and all historical tasks for the current layer, then select a set of candidate tasks $\mathcal{C}$ with the highest similarity scores:} 
\begin{equation*}
    \mathcal{C} = \{ T \in \mathcal{T}_{\text{hist}} \mid \text{Embed-Reuse}^{(l)}(T, T_c) \},
\end{equation*}
% \noindent\textbf{Where:} 
% \begin{itemize}
%     \item $T_c$ represents the current task.
%     \item $\mathcal{T}_{\text{hist}}$ represents all historical tasks recorded in outgoing edges of the current \sysmem node. 
% \end{itemize}
where $T_c$ is the current task, $\mathcal{T}_{\text{hist}}$ represents all historical tasks recorded in outgoing edges of the current \sysmem node.

\FEH{Subsequently, we perform task reranking by comparing each candidate task with the current task. If there exists at least one candidate whose reranking score with the current task is greater than another predefined confidence threshold $\tau_2$, the action at the corresponding node in \sysmem is considered reusable for the current task. This process enables the latent model to make more accurate reuse decisions compared to relying solely on the embedding model.}
The stage-2 (final) reuse decision is expressed as:
\[
\text{Rerank-Reuse}^{(l)} = \begin{cases}
\text{True}, & \text{if } \exists \, T_i \in \mathcal{C} \text{ such that } g(T_i, T_{c},l) \ge \tau_2 \\
\text{False,} & \text{otherwise}
\end{cases}.
\]

\noindent\textbf{Where:} 
\begin{itemize}
    \item $g$ is an \textbf{instruction-aware task reranking model}, and $g(T_i, T_j,l)$ returns a confidence score of the action reuseability between $T_i$ and $T_j$ at $l$-th layer.
\end{itemize}

\subsubsection{Latent Memory Model Training}
We select Qwen-Embedding and Qwen-Reranker~\cite{zhang2025qwen3embeddingadvancingtext} as our base models. All training data are collected through our agent model by recording actual action traces on mobile devices, ensuring behavioral consistency between the Latent Memory Model and the Agent Model.

For the task embedding model, for a given \sysmem layer $l$ and a task $T$, other tasks that share an $l$-step action prefix with $T$ are considered positive samples, while tasks sharing fewer than $l$ steps are considered negative samples. Formally, for $v=f(T_i,l)$, the positive and negative samples are sampled in the following way:

\begin{gather*}
    v^+=f(T_s,l),\; \mathrm{pre}(T_s,T_i) \ge l,\\
    v^-=\left\{ f(T_t,l) \mid  \mathrm{pre}(T_t,T_i) < l\right\},
\end{gather*}

where $\mathrm{pre}(\cdot,\cdot)$ gives the length of the shared action prefix of two tasks, and we calculate the InfoNCE~\cite{oord2019representationlearningcontrastivepredictive} loss with a temperature parameter $\tau$:

$$
\mathcal{L} = \mathbb{E}_v\left[-\ln\frac{e^{S(v, v^{+}) / \tau}}{e^{S(v, v^{+}) / \tau} + \sum_{j=1}^M e^{S(v, v_{j}^{-}) / \tau}}\right],
$$

For the task reranking model, the determination of positive and negative samples is consistent with the embedding model. The positive samples are labeled as ``yes'' and negative samples as ``no'', and we use SFT loss to train the model.

\subsection{Latent Memory Invalidation and Eviction}
Experiences may become obsolete as the environment changes, such as application updates or device replacements. Compared to high-level experiences, low-level experiences (e.g., action coordinates output by the Actor) tend to become outdated more frequently. In the AgentRR system, it is essential to promptly detect and update obsolete experiences to ensure the correctness of subsequent action reuse.

\subsubsection{UI Change Detection} Numerous modern applications utilize dynamic UI components (e.g., the items in search results page change according to user preferences). 
UI change can lead to incorrect action replay in \sysmem when identical actions could yield different results. 
\sysmem leverages OmniParser, a screen parsing tool based on small vision models, to detect and mitigate this anomaly. 
When executing an action with a target UI element (e.g., a click action) for the first time, Tracer invokes OmniParser to extract the target element's bounding box and content. 
Prior to replaying an action during another task's execution, Tracer invokes OmniParser again to verify whether the UI content within the previous bounding box has changed. 
If a UI change is detected, the cache hit is discarded and the cached action is invalidated.

\subsubsection{False Positive of Reuse Decision}
If the latent model outputs a reuse decision of true, but the action cannot actually be reused for the current task, this results in a false positive decision. 
However, false positives in the AgentRR system do not lead to final execution errors. After a false positive action is performed, there will be a significant discrepancy between the current UI state and the previously recorded UI state, causing a failure of UI checking. 
Subsequently, AgentRR will invoke the Decider module again to determine the appropriate action for the current interface, which is typically a back operation.

\subsubsection{LRU Experience Eviction}In addition to experience obsolete, some experiences may rarely be used, resulting in unnecessary consumption of storage resources. To address this, \sysmem implements a Least Recently Used (LRU) eviction policy to prevent out-of-memory issues caused by unused experiences.
When the cache reaches its maximum size, the least recently used experience is removed to make space for new experience.

\subsection{Speculative Replay via Shortcuts}

Task trajectories usually consist of a series of sub-tasks. 
A sub-task is defined as an action sequence following a fixed pattern, e.g., the ``search'' sub-task includes 3 steps: clicking the search bar to activate it, entering the query, and clicking the search button to execute the search. 
As in the aforementioned example, different trajectories of a sub-task can share both prefixes and suffixes (both step 1 and step 3 are reusable).
We refer to frequently recurring sub-tasks as \textbf{shortcuts}. By identifying shortcuts in the tree, \sysmem enables \textbf{speculative replay}.

A shortcut is defined as a set of sub-paths in the \sysmem, denoted as $S$. For any sub-path pair $(p_i,p_j)\in S^2$, $p_i$ and $p_j$ satisfy the following constraints:

\begin{itemize}
    \item They start from the same \sysmem node and have identical lengths.
    \item Their action type sequences before the last action are identical.
    \item Their last actions are identical, denoted as $\mathcal{A}_S^*$.
\end{itemize}

When Tracer reaches a newly created node with no outgoing edges, there are no available experiences to replay. In this scenario, Tracer examines whether the current partially completed task $T$ satisfies the following condition: for some $S$, if $A_S^*$ is the next action for $T$, a sub-path of $T$ can be joined with $S$. If this condition is met, Tracer speculatively selects $A_S^*$ as a candidate for the next action. In this manner, \sysmem expands the scope of experience utilization and enables the discovery of additional replay opportunities.
\section{\sysbench: a DAG-Based Benchmark Framework for Mobile Agents}

Current mobile benchmarks primarily focus on system applications and tasks with clearly defined evaluation criteria. However, in real-world mobile agent scenarios, operations often involve third-party applications and lack explicit verification conditions. As a result, existing mobile agent benchmarks fail to accurately reflect the model’s capabilities in mobile-side scenarios. To address this limitation, we propose an evaluation framework tailored to the manipulation of third-party mobile applications. By leveraging a Directed Acyclic Graph (DAG) to model the complex dependencies and sequential constraints inherent in mobile application tasks, our framework enables a more realistic assessment of model performance in practical on-device agent scenarios.

\begin{figure}[t]
    \centering
    \includegraphics[width=\linewidth]{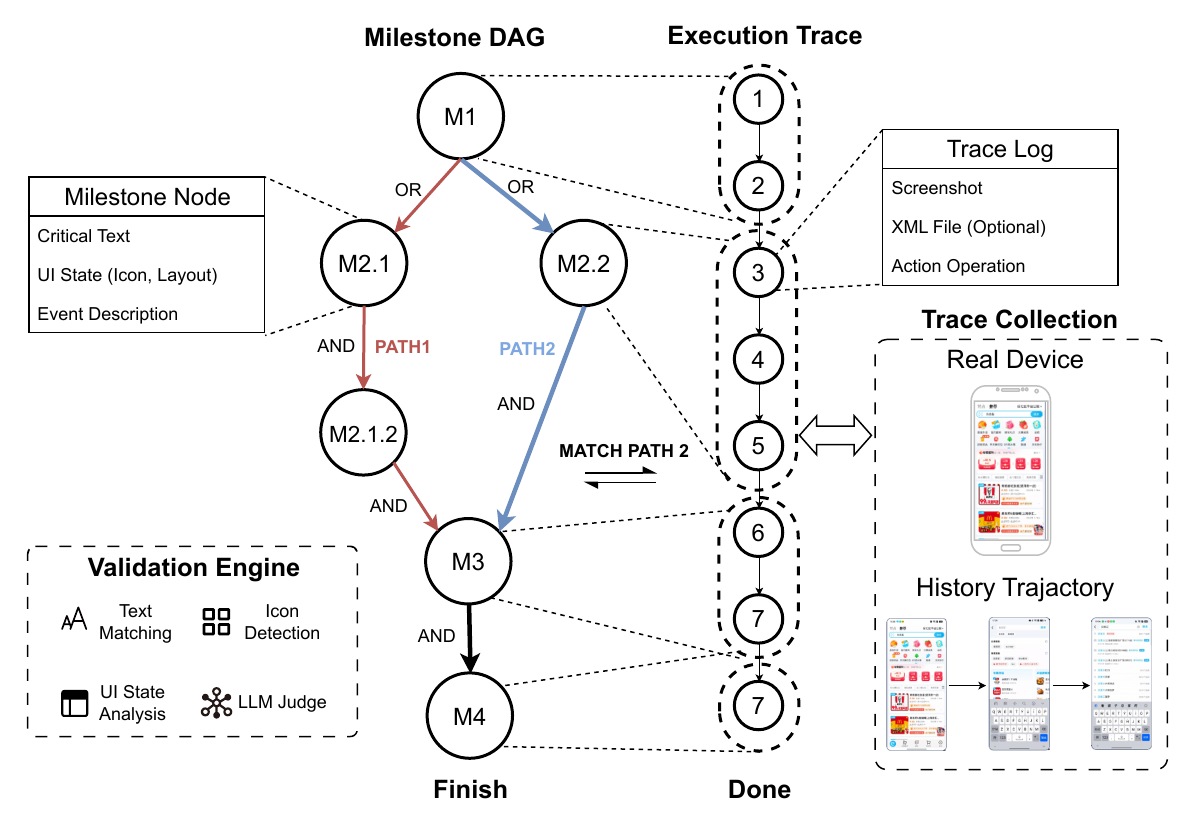}
    \caption{\textbf{The overall architecture of \sysbench}}
    \label{fig:benchmark}
\end{figure}

\subsection{Framework Architecture}
Figure~\ref{fig:benchmark} illustrates our benchmark framework. 
Given the complexity and variability of real-world mobile applications, we define multiple validation methods, diverse task trajectory specifications, 
and both online and offline trajectory collections. 
These design choices aim to accurately capture the agent's capabilities in real-world scenarios. 
The overall architecture of the benchmarking framework is as follows:

\begin{enumerate}
    \item \textbf{Trace Collection}: We categorize agent execution traces into two types: online and offline. Online traces are collected through real interactions between the agent and mobile devices, logging all agent actions as well as UI screenshots from various pages. 
    In contrast, offline traces are obtained by predefining the agent's task execution process and recording information such as action primitives and XML data generated. 
    Additionally, we merge shared steps across different trajectories for the same task. 
    By leveraging offline traces, we can conduct stable evaluations of the agent's performance, effectively eliminating the impact of dynamic environmental variables.
    \item \textbf{DAG-based Task Definition}: Each task is defined by a configuration file that specifies the structure of a DAG. In this graph, nodes represent key milestones in task execution, such as UI states or critical operations, while edges capture the dependencies and action transitions between these milestones.
    \item \textbf{Validation Engine}: The engine parses both the task's DAG configuration and the agent's execution trace. It traverses the DAG alongside the trace, searching within the actual execution trace for milestone events and their dependencies as defined in the DAG.
    \item \textbf{Result Generation}:Based on this traversal, the engine determines whether a fully validated path or a partially validated path exists. Tasks that successfully complete all milestone events defined in the DAG are awarded the full score, while tasks that accomplish only a subset of milestone events may receive partial credit, as specified by user-defined criteria.
\end{enumerate}

In real-world scenarios, application responses are highly dependent on factors such as application version, environment, and interaction patterns. Moreover, there is often a lack of systematic interfaces for quickly verifying whether milestone events have been completed. To address these challenges, we design a multi-level verification mechanism to ensure the accuracy of evaluating agent tasks on the mobile side.

\subsection{Core Validation Capabilities}
The framework supports a hierarchical approach to condition checking for robust and resilient validation.

\textbf{Multi-level Condition Checkers}: Each node in the DAG can be associated with one or more conditions that must be met. The supported checker types include:
\begin{itemize}
    \item \textbf{Text Matching}: Verifies the presence of specific strings or patterns in the UI.
    \item \textbf{Regular Expression Matching}: Allows for more complex text-based validation using regular expressions.
    \item \textbf{UI State Analysis}: Examines the underlying view hierarchy (e.g., via XML dumps) to confirm the state of UI elements.
    \item \textbf{Icon Detection}: Utilizes computer vision techniques to identify the presence of specific icons.
    \item \textbf{OCR Process}: If neither XML files nor the page text is available, we employ OCR techniques to extract and verify text directly from screenshots.
    \item \textbf{MLLM-as-a-Judge}: Employs a multi-modal LLM to make a holistic judgment about the UI state based on the current key frame and contextual information.
\end{itemize}

\textbf{Progressive Escalation Strategy}: The checker starts with the lightest, fastest verification path. When text matching is required, it first parses the XML file and instantly confirms the current page state. Only if the XML source is absent or a plain-text checker is disabled does the system escalate to heavier techniques, like OCR or LLM-based reasoning, to decide whether the milestone has been reached.

\textbf{Multi-path DAG Validation}: The DAG structure inherently supports tasks with multiple valid completion trajectories. For tasks that can be accomplished through different approaches, we specify whether the milestone nodes are dependent on others or can proceed independently in parallel. This allows the DAG representation of a task to include multiple branches and paths. 
To better express the relationships between different paths, we define the \texttt{AND} and \texttt{OR} conjunctions. The \texttt{AND} indicates that all prerequisite milestone events from the preceding nodes must be satisfied before proceeding. In contrast, the \texttt{OR} stipulates that satisfying any single milestone event within the set is sufficient to trigger validation of the subsequent node.
For a given task, if multiple paths can lead to its completion, the task is considered accomplished as long as any one of these paths is successfully validated.

To ensure the correct matching between milestone event nodes defined in the DAG and nodes in the trace, we introduce a synchronized backtracking mechanism for DAG and trace nodes. Specifically, when the checker detects that a certain path in the DAG cannot be validated, it backtracks to the previous branching milestone event and selects an alternative path for verification. Simultaneously, the checker also backtracks within the actual trace until it reaches the milestone node corresponding to the branching point.

\textbf{Dynamic Condition Matching}: \sysbench supports task definition via templates and dynamically extracts conditions from the task description during execution. For example, if the task is ``add milk to the shopping cart'' the validator can automatically identify ``milk'' as the key object and subsequently generate corresponding milestone event nodes in sequence.

\textbf{Icon Recognition}: \sysbench leverages OpenCV~\cite{opencv_library} for icon recognition and supports multi-scale matching. By controlling the similarity threshold, it can robustly identify UI elements, even in the presence of minor variations in size or appearance.

\subsection{Construct the \sysbench Benchmark for Real-world Mobile Agent}
To construct a benchmark that evaluates agent capabilities in realistic mobile environments, we select a range of representative mobile applications and scenarios, encompassing domains such as social networking, music and video, shopping, travel, food delivery, etc. Each test case defines milestone events using a template-based approach, ensuring sufficient generalizability of test cases. Additionally, for each application, we design task cases with varying levels of difficulty, ranging from simple to complex, to accurately assess the capabilities of agent models.

In our actual testing process, we meticulously select tasks that can be correctly executed in the current environment, thereby minimizing test failures caused by task misconfigurations rather than agent performance. For test cases that result in abnormal or failed executions, we further employ manual verification to ensure the accuracy and reliability of the evaluation.

Due to the iteration of applications and the variability of operating environments, 
the absolute scores obtained from \sysbench{} are primarily for reference. 
However, the relative scores among different agent models provide a more accurate reflection of their capabilities in mobile agent scenarios. 
Additionally, we can record the correct action sequences from a given benchmark run as offline traces, 
including multiple valid trajectories for the same task, 
and enable different models to replay these offline traces for evaluation. 
This approach eliminates the influence of dynamic environmental factors, 
thereby making the evaluation results more deterministic.

Nevertheless, evaluating with offline traces may lead to false negatives. 
For example, an agent model could discover a new successful trajectory; 
however, such a correct outcome would be erroneously classified as a failure under offline-trace-based evaluation.
\section{Evaluation}

\subsection{Task Completion Score}
We evaluated the task completion rates of various models in real-world mobile scenarios. Our test suite covers the current widely-used mobile applications in China, as well as common tasks, such as travel, shopping, entertainment and social networking. For each application, we defined a set of test case templates with multiple difficulty levels, where each template consists of milestone events structured in the form of a DAG. Prior to formal testing, we generated three real-world test cases for each type of template. All test cases are defined within our MobiFlow benchmark framework.

% --- 将三张图横向排列的完整代码块 ---
\begin{figure}[t!]
    \centering % 整体居中
    % --- 图 1 ---
    \begin{subfigure}[b]{0.32\textwidth}
        \centering
        \includegraphics[width=\linewidth]{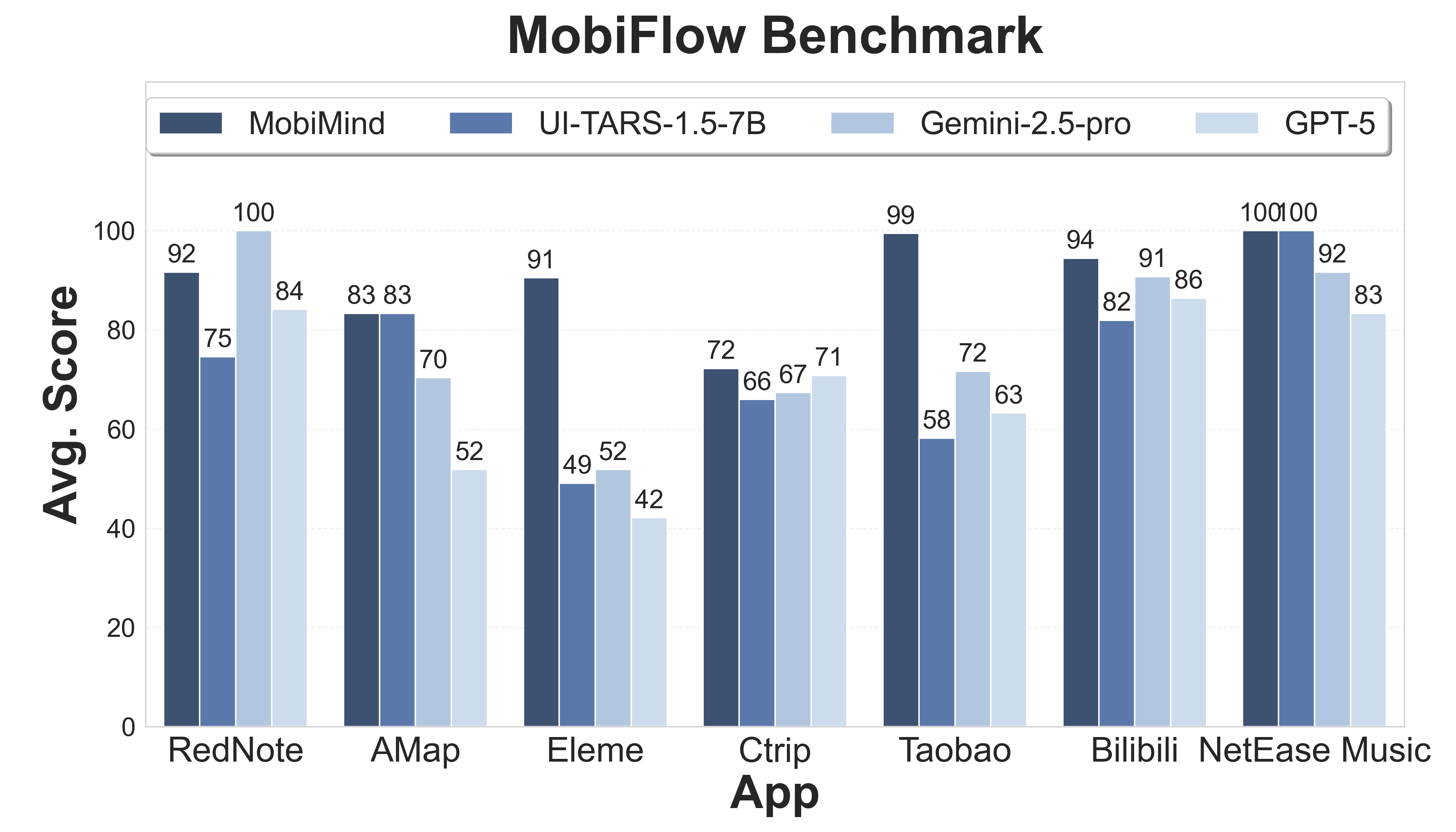}
        \caption{Overall Average Score}
        \label{fig:avg_score}
    \end{subfigure}% 
    \hfill % 
    \begin{subfigure}[b]{0.32\textwidth}
        \centering
        \includegraphics[width=\linewidth]{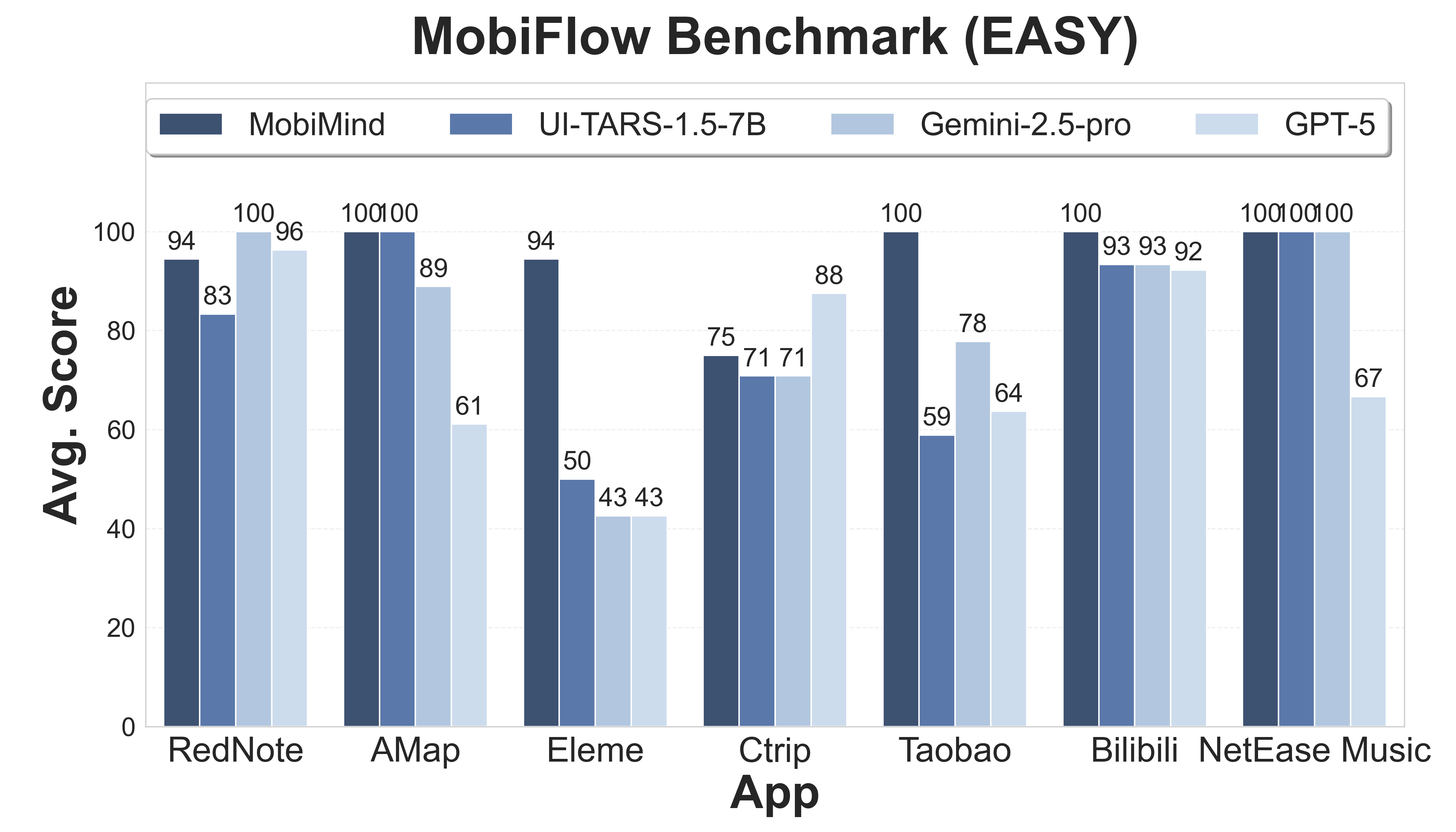}
        \caption{Average Easy Task Score}
        \label{fig:avg_score_easy}
    \end{subfigure}% 
    \hfill % 
    \begin{subfigure}[b]{0.32\textwidth}
        \centering
        \includegraphics[width=\linewidth]{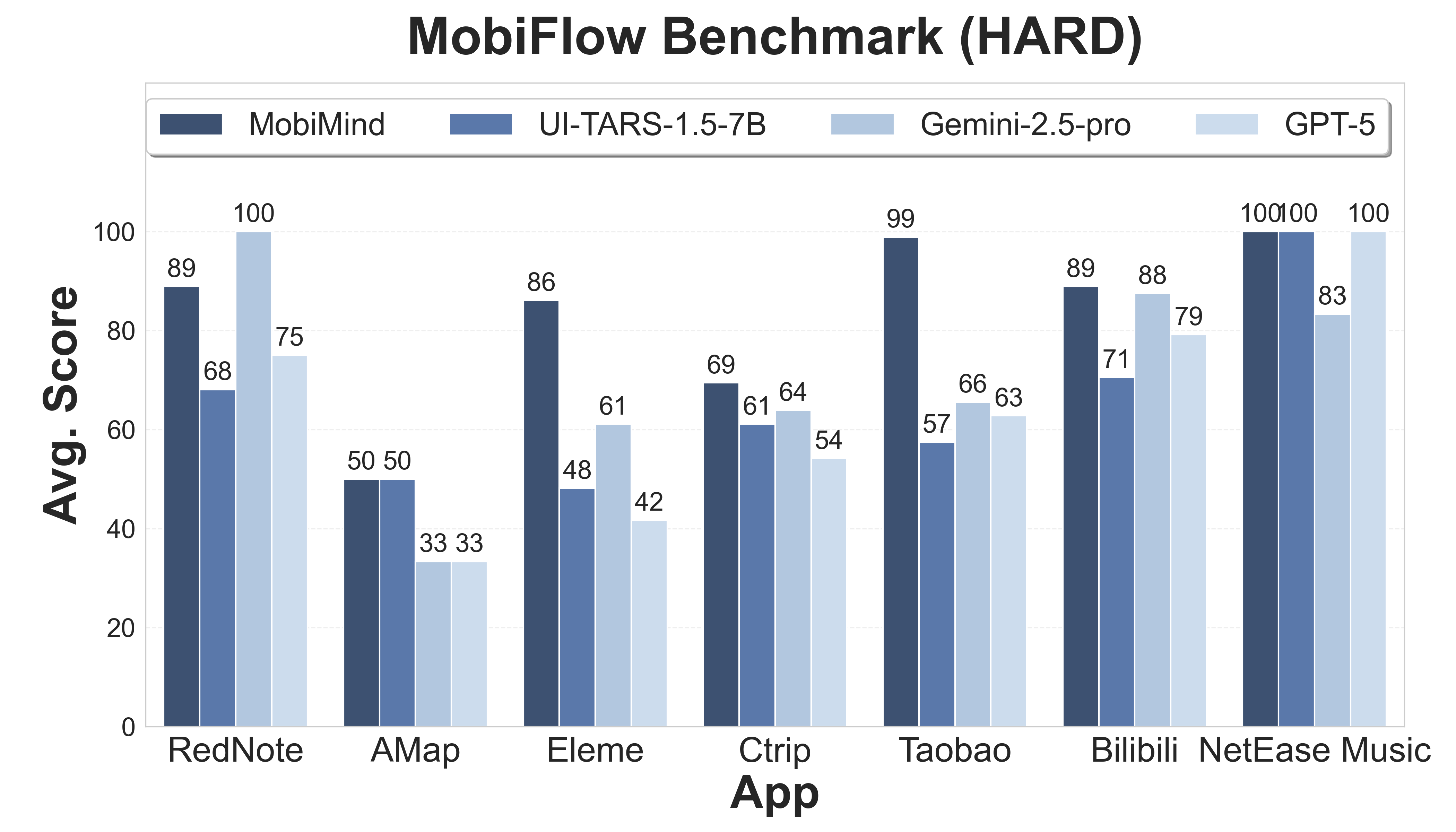}
        \caption{Average Hard Task Score}
        \label{fig:avg_score_hard}
    \end{subfigure}
    \caption{\textbf{Average Task Completion Scores of Different Agent Models Under Realistic Workloads (MobiFlow):} The performance is shown for (a) all tasks combined; (b) easy tasks; (c) and hard tasks.}
    \label{fig:all_scores_side_by_side}
\end{figure}

Figure~\ref{fig:all_scores_side_by_side} presents a comparison of task completion rates between MobiAgent (MobiMind-Decider-7B + MobiMind-Grounder-3B) and several other VLM models, including the general-purpose LLMs (GPT-5, Gemini-2.5-pro) and the state-of-the-art GUI Agent model UI-TARS-1.5-7B. To ensure the accuracy of our evaluation, we manually verified all failed tasks. For cases caused by app anomalies, network interruptions, or non-existent tasks, we conducted repeated testing. In addition, we imposed a termination penalty on tasks where the agent could not complete or exit properly.

MobiAgent achieves the highest task completion rates across the majority of applications. In complex tasks such as shopping and food delivery, MobiAgent demonstrates superior performance in task understanding and decomposition, instruction following, and handling of exceptional cases.
During execution, GPT and Gemini tend to input object descriptions, types, and details such as departure time and locations, directly into the search box. Because some apps support AI-powered search capabilities, these models can bypass most complex interactive steps, resulting in relatively higher task completion rates.
However, such coarse-grained operation strategies lead to a significant drop in completion rates when faced with applications that do not support AI search.

Furthermore, we observed that existing agent models still suffer from task non-termination issues, such as the infinite repetition of certain actions. Specifically, GPT failed to properly terminate tasks in 11 application categories, while Gemini exhibited similar problems in 3 categories. In contrast, MobiAgent consistently achieved correct task termination across all evaluated scenarios.

\subsection{Accelerating the Mobile Agent using AgentRR}

In addition to evaluating the task completion rates of agents, we also measure the speedup achieved by the AgentRR acceleration framework across various scenarios. By leveraging \sysmem and a latent memory model, AgentRR can rapidly reuse historical agent trajectories, thereby reducing the inference overhead of VLMs. Consequently, the acceleration ratio of AgentRR in real-world mobile scenarios is positively correlated with the proportion of nodes in \sysmem that can be reused.

To simulate different user behaviors on mobile devices, we constructed two types of user task distributions: a uniform distribution and a power law distribution. In the uniform distribution, users evenly utilize various application functionalities and randomly choose task parameters (e.g., search objects). In the power law distribution, 80\% of user requests are concentrated within 20\% of tasks, which more closely aligns with real-world user habits.

\begin{figure}[htp]
    \centering
    \includegraphics[width=\linewidth]{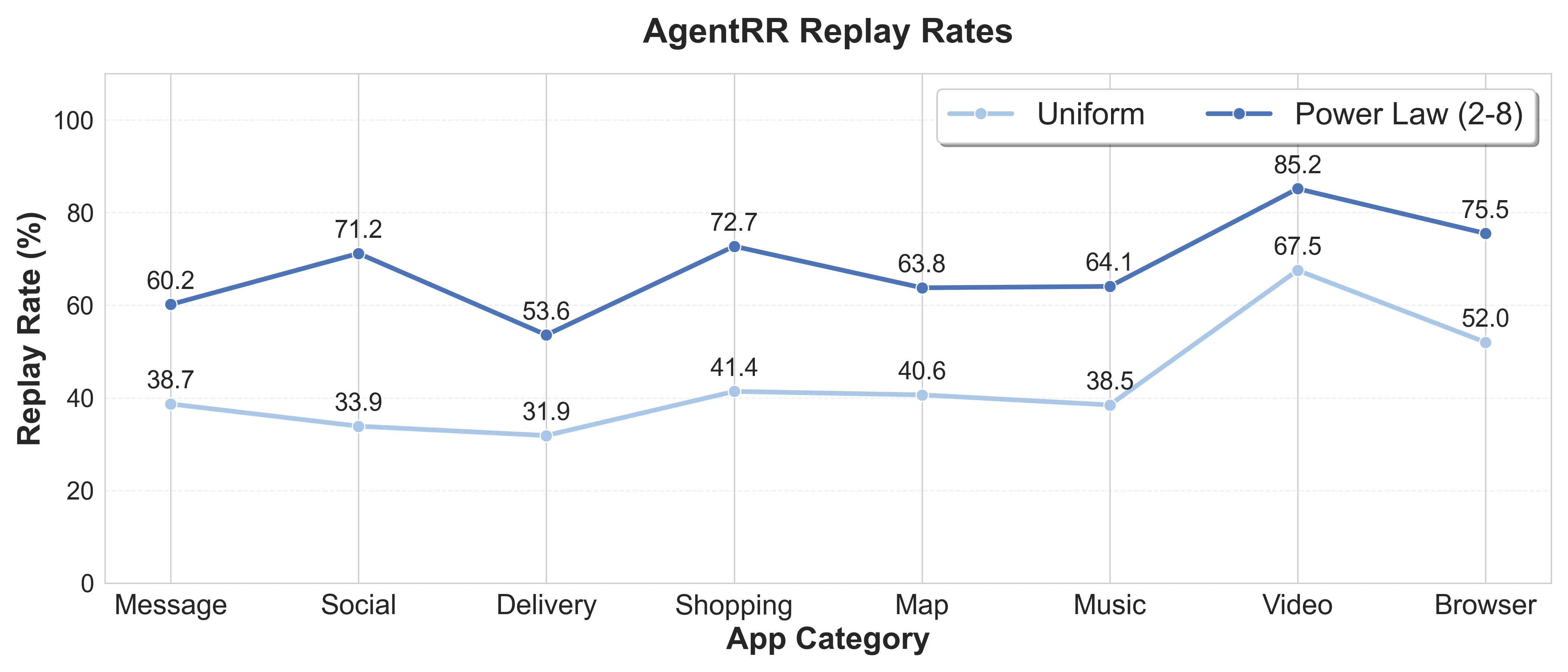}
    \caption{\textbf{Action Replay Rate with AgentRR under Different Distributions.}}
    \label{fig:replay}
\end{figure}

As Figure~\ref{fig:replay} shows, under the uniform distribution, AgentRR achieves an action replay rate of 30\%--60\%. Under the power law distribution, the action replay rate increases to 60\%--85\%. Moreover, owing to the accuracy of the latent model, the correctness of action replay exceeds 99\% in our test cases.

In the real-world deployment scenarios, the performance advantage of AgentRR becomes even more pronounced, as the latent memory model can be deployed directly on edge devices, thereby reducing network communication overhead. In complex task scenarios such as food delivery and online shopping, AgentRR achieves an average performance improvement of 2 to 3 times compared to baseline approaches.
\section{Conclusion}
In this paper, we address the challenges of building, evaluating and deploying mobile agent by introducing a comprehensive, full-stack solution encompassing the MobiMind model series, the AgentRR acceleration framework, and the \sysbench evaluation benchmark. Experiments show that MobiAgent achieves state-of-the-art task completion rates that surpass both general-purpose models like GPT-5 and Gemini-2.5 Pro, and other specialized mobile agents in the \sysbench benchmark. Moreover, The experience replay mechanism of AgentRR framework delivers a 2-3x optimization on agent's task completion latency in real-world scenarios. These results demonstrate the effectiveness of our approach in creating customizable and practical mobile agents.

\bibliographystyle{ACM-Reference-Format}
\bibliography{ref}

%%% -*-BibTeX-*-
%%% Do NOT edit. File created by BibTeX with style
%%% ACM-Reference-Format-Journals [18-Jan-2012].

\begin{thebibliography}{34}

%%% ====================================================================
%%% NOTE TO THE USER: you can override these defaults by providing
%%% customized versions of any of these macros before the \bibliography
%%% command.  Each of them MUST provide its own final punctuation,
%%% except for \shownote{}, \showDOI{}, and \showURL{}.  The latter two
%%% do not use final punctuation, in order to avoid confusing it with
%%% the Web address.
%%%
%%% To suppress output of a particular field, define its macro to expand
%%% to an empty string, or better, \unskip, like this:
%%%
%%% \newcommand{\showDOI}[1]{\unskip}   % LaTeX syntax
%%%
%%% \def \showDOI #1{\unskip}           % plain TeX syntax
%%%
%%% ====================================================================

\ifx \showCODEN    \undefined \def \showCODEN     #1{\unskip}     \fi
\ifx \showDOI      \undefined \def \showDOI       #1{#1}\fi
\ifx \showISBNx    \undefined \def \showISBNx     #1{\unskip}     \fi
\ifx \showISBNxiii \undefined \def \showISBNxiii  #1{\unskip}     \fi
\ifx \showISSN     \undefined \def \showISSN      #1{\unskip}     \fi
\ifx \showLCCN     \undefined \def \showLCCN      #1{\unskip}     \fi
\ifx \shownote     \undefined \def \shownote      #1{#1}          \fi
\ifx \showarticletitle \undefined \def \showarticletitle #1{#1}   \fi
\ifx \showURL      \undefined \def \showURL       {\relax}        \fi
% The following commands are used for tagged output and should be
% invisible to TeX
\providecommand\bibfield[2]{#2}
\providecommand\bibinfo[2]{#2}
\providecommand\natexlab[1]{#1}
\providecommand\showeprint[2][]{arXiv:#2}

\bibitem[Bai et~al\mbox{.}(2025)]%
        {bai2025qwen25vltechnicalreport}
\bibfield{author}{\bibinfo{person}{Shuai Bai}, \bibinfo{person}{Keqin Chen}, \bibinfo{person}{Xuejing Liu}, \bibinfo{person}{Jialin Wang}, \bibinfo{person}{Wenbin Ge}, \bibinfo{person}{Sibo Song}, \bibinfo{person}{Kai Dang}, \bibinfo{person}{Peng Wang}, \bibinfo{person}{Shijie Wang}, \bibinfo{person}{Jun Tang}, \bibinfo{person}{Humen Zhong}, \bibinfo{person}{Yuanzhi Zhu}, \bibinfo{person}{Mingkun Yang}, \bibinfo{person}{Zhaohai Li}, \bibinfo{person}{Jianqiang Wan}, \bibinfo{person}{Pengfei Wang}, \bibinfo{person}{Wei Ding}, \bibinfo{person}{Zheren Fu}, \bibinfo{person}{Yiheng Xu}, \bibinfo{person}{Jiabo Ye}, \bibinfo{person}{Xi Zhang}, \bibinfo{person}{Tianbao Xie}, \bibinfo{person}{Zesen Cheng}, \bibinfo{person}{Hang Zhang}, \bibinfo{person}{Zhibo Yang}, \bibinfo{person}{Haiyang Xu}, {and} \bibinfo{person}{Junyang Lin}.} \bibinfo{year}{2025}\natexlab{}.
\newblock \bibinfo{title}{Qwen2.5-VL Technical Report}.
\newblock
\newblock
\showeprint[arxiv]{2502.13923}~[cs.CV]
\urldef\tempurl%
\url{https://arxiv.org/abs/2502.13923}
\showURL{%
\tempurl}


\bibitem[Bengio et~al\mbox{.}(2009)]%
        {10.1145/1553374.1553380}
\bibfield{author}{\bibinfo{person}{Yoshua Bengio}, \bibinfo{person}{J\'{e}r\^{o}me Louradour}, \bibinfo{person}{Ronan Collobert}, {and} \bibinfo{person}{Jason Weston}.} \bibinfo{year}{2009}\natexlab{}.
\newblock \showarticletitle{Curriculum learning}. In \bibinfo{booktitle}{\emph{Proceedings of the 26th Annual International Conference on Machine Learning}} (Montreal, Quebec, Canada) \emph{(\bibinfo{series}{ICML '09})}. \bibinfo{publisher}{Association for Computing Machinery}, \bibinfo{address}{New York, NY, USA}, \bibinfo{pages}{41–48}.
\newblock
\showISBNx{9781605585161}
\urldef\tempurl%
\url{https://doi.org/10.1145/1553374.1553380}
\showDOI{\tempurl}


\bibitem[Bradski(2000)]%
        {opencv_library}
\bibfield{author}{\bibinfo{person}{G. Bradski}.} \bibinfo{year}{2000}\natexlab{}.
\newblock \showarticletitle{{The OpenCV Library}}.
\newblock \bibinfo{journal}{\emph{Dr. Dobb's Journal of Software Tools}} (\bibinfo{year}{2000}).
\newblock


\bibitem[Chai et~al\mbox{.}(2025)]%
        {chai2025a3androidagentarena}
\bibfield{author}{\bibinfo{person}{Yuxiang Chai}, \bibinfo{person}{Hanhao Li}, \bibinfo{person}{Jiayu Zhang}, \bibinfo{person}{Liang Liu}, \bibinfo{person}{Guangyi Liu}, \bibinfo{person}{Guozhi Wang}, \bibinfo{person}{Shuai Ren}, \bibinfo{person}{Siyuan Huang}, {and} \bibinfo{person}{Hongsheng Li}.} \bibinfo{year}{2025}\natexlab{}.
\newblock \bibinfo{title}{A3: Android Agent Arena for Mobile GUI Agents}.
\newblock
\newblock
\showeprint[arxiv]{2501.01149}~[cs.AI]
\urldef\tempurl%
\url{https://arxiv.org/abs/2501.01149}
\showURL{%
\tempurl}


\bibitem[Cheng et~al\mbox{.}(2024)]%
        {cheng2024seeclickharnessingguigrounding}
\bibfield{author}{\bibinfo{person}{Kanzhi Cheng}, \bibinfo{person}{Qiushi Sun}, \bibinfo{person}{Yougang Chu}, \bibinfo{person}{Fangzhi Xu}, \bibinfo{person}{Yantao Li}, \bibinfo{person}{Jianbing Zhang}, {and} \bibinfo{person}{Zhiyong Wu}.} \bibinfo{year}{2024}\natexlab{}.
\newblock \bibinfo{title}{SeeClick: Harnessing GUI Grounding for Advanced Visual GUI Agents}.
\newblock
\newblock
\showeprint[arxiv]{2401.10935}~[cs.HC]
\urldef\tempurl%
\url{https://arxiv.org/abs/2401.10935}
\showURL{%
\tempurl}


\bibitem[Feng et~al\mbox{.}(2025)]%
        {feng2025experiencepracticellmagents}
\bibfield{author}{\bibinfo{person}{Erhu Feng}, \bibinfo{person}{Wenbo Zhou}, \bibinfo{person}{Zibin Liu}, \bibinfo{person}{Le Chen}, \bibinfo{person}{Yunpeng Dong}, \bibinfo{person}{Cheng Zhang}, \bibinfo{person}{Yisheng Zhao}, \bibinfo{person}{Dong Du}, \bibinfo{person}{Zhichao Hua}, \bibinfo{person}{Yubin Xia}, {and} \bibinfo{person}{Haibo Chen}.} \bibinfo{year}{2025}\natexlab{}.
\newblock \bibinfo{title}{Get Experience from Practice: LLM Agents with Record \& Replay}.
\newblock
\newblock
\showeprint[arxiv]{2505.17716}~[cs.LG]
\urldef\tempurl%
\url{https://arxiv.org/abs/2505.17716}
\showURL{%
\tempurl}


\bibitem[Google(2025)]%
        {gemini-2.5-pro}
\bibfield{author}{\bibinfo{person}{Google}.} \bibinfo{year}{2025}\natexlab{}.
\newblock \bibinfo{title}{Gemini 2.5 Pro Best for coding and highly complex tasks}.
\newblock
\newblock
\urldef\tempurl%
\url{https://deepmind.google/models/gemini/pro/}
\showURL{%
\tempurl}


\bibitem[Gou et~al\mbox{.}(2025)]%
        {gou2024uground}
\bibfield{author}{\bibinfo{person}{Boyu Gou}, \bibinfo{person}{Ruohan Wang}, \bibinfo{person}{Boyuan Zheng}, \bibinfo{person}{Yanan Xie}, \bibinfo{person}{Cheng Chang}, \bibinfo{person}{Yiheng Shu}, \bibinfo{person}{Huan Sun}, {and} \bibinfo{person}{Yu Su}.} \bibinfo{year}{2025}\natexlab{}.
\newblock \showarticletitle{Navigating the Digital World as Humans Do: Universal Visual Grounding for {GUI} Agents}. In \bibinfo{booktitle}{\emph{The Thirteenth International Conference on Learning Representations}}.
\newblock
\urldef\tempurl%
\url{https://openreview.net/forum?id=kxnoqaisCT}
\showURL{%
\tempurl}


\bibitem[Hong et~al\mbox{.}(2024)]%
        {hong2024cogagentvisuallanguagemodel}
\bibfield{author}{\bibinfo{person}{Wenyi Hong}, \bibinfo{person}{Weihan Wang}, \bibinfo{person}{Qingsong Lv}, \bibinfo{person}{Jiazheng Xu}, \bibinfo{person}{Wenmeng Yu}, \bibinfo{person}{Junhui Ji}, \bibinfo{person}{Yan Wang}, \bibinfo{person}{Zihan Wang}, \bibinfo{person}{Yuxuan Zhang}, \bibinfo{person}{Juanzi Li}, \bibinfo{person}{Bin Xu}, \bibinfo{person}{Yuxiao Dong}, \bibinfo{person}{Ming Ding}, {and} \bibinfo{person}{Jie Tang}.} \bibinfo{year}{2024}\natexlab{}.
\newblock \bibinfo{title}{CogAgent: A Visual Language Model for GUI Agents}.
\newblock
\newblock
\showeprint[arxiv]{2312.08914}~[cs.CV]
\urldef\tempurl%
\url{https://arxiv.org/abs/2312.08914}
\showURL{%
\tempurl}


\bibitem[Kapoor et~al\mbox{.}(2024)]%
        {kapoor2024omniactdatasetbenchmarkenabling}
\bibfield{author}{\bibinfo{person}{Raghav Kapoor}, \bibinfo{person}{Yash~Parag Butala}, \bibinfo{person}{Melisa Russak}, \bibinfo{person}{Jing~Yu Koh}, \bibinfo{person}{Kiran Kamble}, \bibinfo{person}{Waseem Alshikh}, {and} \bibinfo{person}{Ruslan Salakhutdinov}.} \bibinfo{year}{2024}\natexlab{}.
\newblock \bibinfo{title}{OmniACT: A Dataset and Benchmark for Enabling Multimodal Generalist Autonomous Agents for Desktop and Web}.
\newblock
\newblock
\showeprint[arxiv]{2402.17553}~[cs.AI]
\urldef\tempurl%
\url{https://arxiv.org/abs/2402.17553}
\showURL{%
\tempurl}


\bibitem[Lee et~al\mbox{.}(2024b)]%
        {lee2024benchmarking}
\bibfield{author}{\bibinfo{person}{Juyong Lee}, \bibinfo{person}{Taywon Min}, \bibinfo{person}{Minyong An}, \bibinfo{person}{Dongyoon Hahm}, \bibinfo{person}{Haeone Lee}, \bibinfo{person}{Changyeon Kim}, {and} \bibinfo{person}{Kimin Lee}.} \bibinfo{year}{2024}\natexlab{b}.
\newblock \showarticletitle{Benchmarking Mobile Device Control Agents Across Diverse Configurations}.
\newblock \bibinfo{journal}{\emph{arXiv preprint arXiv:2404.16660}} (\bibinfo{year}{2024}).
\newblock


\bibitem[Lee et~al\mbox{.}(2024a)]%
        {10.1145/3636534.3690682}
\bibfield{author}{\bibinfo{person}{Sunjae Lee}, \bibinfo{person}{Junyoung Choi}, \bibinfo{person}{Jungjae Lee}, \bibinfo{person}{Munim~Hasan Wasi}, \bibinfo{person}{Hojun Choi}, \bibinfo{person}{Steve Ko}, \bibinfo{person}{Sangeun Oh}, {and} \bibinfo{person}{Insik Shin}.} \bibinfo{year}{2024}\natexlab{a}.
\newblock \showarticletitle{MobileGPT: Augmenting LLM with Human-like App Memory for Mobile Task Automation}. In \bibinfo{booktitle}{\emph{Proceedings of the 30th Annual International Conference on Mobile Computing and Networking}} (Washington D.C., DC, USA) \emph{(\bibinfo{series}{ACM MobiCom '24})}. \bibinfo{publisher}{Association for Computing Machinery}, \bibinfo{address}{New York, NY, USA}, \bibinfo{pages}{1119–1133}.
\newblock
\showISBNx{9798400704895}
\urldef\tempurl%
\url{https://doi.org/10.1145/3636534.3690682}
\showDOI{\tempurl}


\bibitem[Li et~al\mbox{.}(2025)]%
        {li2025screenspotproguigroundingprofessional}
\bibfield{author}{\bibinfo{person}{Kaixin Li}, \bibinfo{person}{Ziyang Meng}, \bibinfo{person}{Hongzhan Lin}, \bibinfo{person}{Ziyang Luo}, \bibinfo{person}{Yuchen Tian}, \bibinfo{person}{Jing Ma}, \bibinfo{person}{Zhiyong Huang}, {and} \bibinfo{person}{Tat-Seng Chua}.} \bibinfo{year}{2025}\natexlab{}.
\newblock \bibinfo{title}{ScreenSpot-Pro: GUI Grounding for Professional High-Resolution Computer Use}.
\newblock
\newblock
\showeprint[arxiv]{2504.07981}~[cs.CV]
\urldef\tempurl%
\url{https://arxiv.org/abs/2504.07981}
\showURL{%
\tempurl}


\bibitem[Li et~al\mbox{.}(2024)]%
        {li2024effectsdatascaleui}
\bibfield{author}{\bibinfo{person}{Wei Li}, \bibinfo{person}{William Bishop}, \bibinfo{person}{Alice Li}, \bibinfo{person}{Chris Rawles}, \bibinfo{person}{Folawiyo Campbell-Ajala}, \bibinfo{person}{Divya Tyamagundlu}, {and} \bibinfo{person}{Oriana Riva}.} \bibinfo{year}{2024}\natexlab{}.
\newblock \bibinfo{title}{On the Effects of Data Scale on UI Control Agents}.
\newblock
\newblock
\showeprint[arxiv]{2406.03679}~[cs.AI]
\urldef\tempurl%
\url{https://arxiv.org/abs/2406.03679}
\showURL{%
\tempurl}


\bibitem[Lin et~al\mbox{.}(2024)]%
        {lin2024showuivisionlanguageactionmodelgui}
\bibfield{author}{\bibinfo{person}{Kevin~Qinghong Lin}, \bibinfo{person}{Linjie Li}, \bibinfo{person}{Difei Gao}, \bibinfo{person}{Zhengyuan Yang}, \bibinfo{person}{Shiwei Wu}, \bibinfo{person}{Zechen Bai}, \bibinfo{person}{Weixian Lei}, \bibinfo{person}{Lijuan Wang}, {and} \bibinfo{person}{Mike~Zheng Shou}.} \bibinfo{year}{2024}\natexlab{}.
\newblock \bibinfo{title}{ShowUI: One Vision-Language-Action Model for GUI Visual Agent}.
\newblock
\newblock
\showeprint[arxiv]{2411.17465}~[cs.CV]
\urldef\tempurl%
\url{https://arxiv.org/abs/2411.17465}
\showURL{%
\tempurl}


\bibitem[Lu et~al\mbox{.}(2025)]%
        {lu2025guiodysseycomprehensivedatasetcrossapp}
\bibfield{author}{\bibinfo{person}{Quanfeng Lu}, \bibinfo{person}{Wenqi Shao}, \bibinfo{person}{Zitao Liu}, \bibinfo{person}{Lingxiao Du}, \bibinfo{person}{Fanqing Meng}, \bibinfo{person}{Boxuan Li}, \bibinfo{person}{Botong Chen}, \bibinfo{person}{Siyuan Huang}, \bibinfo{person}{Kaipeng Zhang}, {and} \bibinfo{person}{Ping Luo}.} \bibinfo{year}{2025}\natexlab{}.
\newblock \bibinfo{title}{GUIOdyssey: A Comprehensive Dataset for Cross-App GUI Navigation on Mobile Devices}.
\newblock
\newblock
\showeprint[arxiv]{2406.08451}~[cs.CV]
\urldef\tempurl%
\url{https://arxiv.org/abs/2406.08451}
\showURL{%
\tempurl}


\bibitem[Lu et~al\mbox{.}(2024)]%
        {lu2024omniparserpurevisionbased}
\bibfield{author}{\bibinfo{person}{Yadong Lu}, \bibinfo{person}{Jianwei Yang}, \bibinfo{person}{Yelong Shen}, {and} \bibinfo{person}{Ahmed Awadallah}.} \bibinfo{year}{2024}\natexlab{}.
\newblock \bibinfo{title}{OmniParser for Pure Vision Based GUI Agent}.
\newblock
\newblock
\showeprint[arxiv]{2408.00203}~[cs.CV]
\urldef\tempurl%
\url{https://arxiv.org/abs/2408.00203}
\showURL{%
\tempurl}


\bibitem[OpenAI(2025)]%
        {GPT-5}
\bibfield{author}{\bibinfo{person}{OpenAI}.} \bibinfo{year}{2025}\natexlab{}.
\newblock \bibinfo{title}{GPT-5 is here}.
\newblock
\newblock
\urldef\tempurl%
\url{https://openai.com/gpt-5/}
\showURL{%
\tempurl}


\bibitem[Pan et~al\mbox{.}(2024)]%
        {pan2024autonomous}
\bibfield{author}{\bibinfo{person}{Jiayi Pan}, \bibinfo{person}{Yichi Zhang}, \bibinfo{person}{Nicholas Tomlin}, \bibinfo{person}{Yifei Zhou}, \bibinfo{person}{Sergey Levine}, {and} \bibinfo{person}{Alane Suhr}.} \bibinfo{year}{2024}\natexlab{}.
\newblock \bibinfo{title}{Autonomous Evaluation and Refinement of Digital Agents}.
\newblock
\newblock
\showeprint[arxiv]{2404.06474}~[cs.AI]


\bibitem[Qin et~al\mbox{.}(2025)]%
        {qin2025uitarspioneeringautomatedgui}
\bibfield{author}{\bibinfo{person}{Yujia Qin}, \bibinfo{person}{Yining Ye}, \bibinfo{person}{Junjie Fang}, \bibinfo{person}{Haoming Wang}, \bibinfo{person}{Shihao Liang}, \bibinfo{person}{Shizuo Tian}, \bibinfo{person}{Junda Zhang}, \bibinfo{person}{Jiahao Li}, \bibinfo{person}{Yunxin Li}, \bibinfo{person}{Shijue Huang}, \bibinfo{person}{Wanjun Zhong}, \bibinfo{person}{Kuanye Li}, \bibinfo{person}{Jiale Yang}, \bibinfo{person}{Yu Miao}, \bibinfo{person}{Woyu Lin}, \bibinfo{person}{Longxiang Liu}, \bibinfo{person}{Xu Jiang}, \bibinfo{person}{Qianli Ma}, \bibinfo{person}{Jingyu Li}, \bibinfo{person}{Xiaojun Xiao}, \bibinfo{person}{Kai Cai}, \bibinfo{person}{Chuang Li}, \bibinfo{person}{Yaowei Zheng}, \bibinfo{person}{Chaolin Jin}, \bibinfo{person}{Chen Li}, \bibinfo{person}{Xiao Zhou}, \bibinfo{person}{Minchao Wang}, \bibinfo{person}{Haoli Chen}, \bibinfo{person}{Zhaojian Li}, \bibinfo{person}{Haihua Yang}, \bibinfo{person}{Haifeng Liu}, \bibinfo{person}{Feng Lin}, \bibinfo{person}{Tao Peng}, \bibinfo{person}{Xin Liu}, {and} \bibinfo{person}{Guang Shi}.} \bibinfo{year}{2025}\natexlab{}.
\newblock \bibinfo{title}{UI-TARS: Pioneering Automated GUI Interaction with Native Agents}.
\newblock
\newblock
\showeprint[arxiv]{2501.12326}~[cs.AI]
\urldef\tempurl%
\url{https://arxiv.org/abs/2501.12326}
\showURL{%
\tempurl}


\bibitem[Rawles et~al\mbox{.}(2025)]%
        {rawles2025androidworlddynamicbenchmarkingenvironment}
\bibfield{author}{\bibinfo{person}{Christopher Rawles}, \bibinfo{person}{Sarah Clinckemaillie}, \bibinfo{person}{Yifan Chang}, \bibinfo{person}{Jonathan Waltz}, \bibinfo{person}{Gabrielle Lau}, \bibinfo{person}{Marybeth Fair}, \bibinfo{person}{Alice Li}, \bibinfo{person}{William Bishop}, \bibinfo{person}{Wei Li}, \bibinfo{person}{Folawiyo Campbell-Ajala}, \bibinfo{person}{Daniel Toyama}, \bibinfo{person}{Robert Berry}, \bibinfo{person}{Divya Tyamagundlu}, \bibinfo{person}{Timothy Lillicrap}, {and} \bibinfo{person}{Oriana Riva}.} \bibinfo{year}{2025}\natexlab{}.
\newblock \bibinfo{title}{AndroidWorld: A Dynamic Benchmarking Environment for Autonomous Agents}.
\newblock
\newblock
\showeprint[arxiv]{2405.14573}~[cs.AI]
\urldef\tempurl%
\url{https://arxiv.org/abs/2405.14573}
\showURL{%
\tempurl}


\bibitem[Rawles et~al\mbox{.}(2023)]%
        {10.5555/3666122.3668731}
\bibfield{author}{\bibinfo{person}{Christopher Rawles}, \bibinfo{person}{Alice Li}, \bibinfo{person}{Daniel Rodriguez}, \bibinfo{person}{Oriana Riva}, {and} \bibinfo{person}{Timothy Lillicrap}.} \bibinfo{year}{2023}\natexlab{}.
\newblock \showarticletitle{Android in the wild: a large-scale dataset for android device control}. In \bibinfo{booktitle}{\emph{Proceedings of the 37th International Conference on Neural Information Processing Systems}} (New Orleans, LA, USA) \emph{(\bibinfo{series}{NIPS '23})}. \bibinfo{publisher}{Curran Associates Inc.}, \bibinfo{address}{Red Hook, NY, USA}, Article \bibinfo{articleno}{2609}, \bibinfo{numpages}{21}~pages.
\newblock


\bibitem[Shao et~al\mbox{.}(2024)]%
        {shao2024deepseekmathpushinglimitsmathematical}
\bibfield{author}{\bibinfo{person}{Zhihong Shao}, \bibinfo{person}{Peiyi Wang}, \bibinfo{person}{Qihao Zhu}, \bibinfo{person}{Runxin Xu}, \bibinfo{person}{Junxiao Song}, \bibinfo{person}{Xiao Bi}, \bibinfo{person}{Haowei Zhang}, \bibinfo{person}{Mingchuan Zhang}, \bibinfo{person}{Y.~K. Li}, \bibinfo{person}{Y. Wu}, {and} \bibinfo{person}{Daya Guo}.} \bibinfo{year}{2024}\natexlab{}.
\newblock \bibinfo{title}{DeepSeekMath: Pushing the Limits of Mathematical Reasoning in Open Language Models}.
\newblock
\newblock
\showeprint[arxiv]{2402.03300}~[cs.CL]
\urldef\tempurl%
\url{https://arxiv.org/abs/2402.03300}
\showURL{%
\tempurl}


\bibitem[van~den Oord et~al\mbox{.}(2019)]%
        {oord2019representationlearningcontrastivepredictive}
\bibfield{author}{\bibinfo{person}{Aaron van~den Oord}, \bibinfo{person}{Yazhe Li}, {and} \bibinfo{person}{Oriol Vinyals}.} \bibinfo{year}{2019}\natexlab{}.
\newblock \bibinfo{title}{Representation Learning with Contrastive Predictive Coding}.
\newblock
\newblock
\showeprint[arxiv]{1807.03748}~[cs.LG]
\urldef\tempurl%
\url{https://arxiv.org/abs/1807.03748}
\showURL{%
\tempurl}


\bibitem[Wang et~al\mbox{.}(2024)]%
        {wang2024mobileagentv2mobiledeviceoperation}
\bibfield{author}{\bibinfo{person}{Junyang Wang}, \bibinfo{person}{Haiyang Xu}, \bibinfo{person}{Haitao Jia}, \bibinfo{person}{Xi Zhang}, \bibinfo{person}{Ming Yan}, \bibinfo{person}{Weizhou Shen}, \bibinfo{person}{Ji Zhang}, \bibinfo{person}{Fei Huang}, {and} \bibinfo{person}{Jitao Sang}.} \bibinfo{year}{2024}\natexlab{}.
\newblock \bibinfo{title}{Mobile-Agent-v2: Mobile Device Operation Assistant with Effective Navigation via Multi-Agent Collaboration}.
\newblock
\newblock
\showeprint[arxiv]{2406.01014}~[cs.CL]
\urldef\tempurl%
\url{https://arxiv.org/abs/2406.01014}
\showURL{%
\tempurl}


\bibitem[Wen et~al\mbox{.}(2024)]%
        {10.1145/3636534.3649379}
\bibfield{author}{\bibinfo{person}{Hao Wen}, \bibinfo{person}{Yuanchun Li}, \bibinfo{person}{Guohong Liu}, \bibinfo{person}{Shanhui Zhao}, \bibinfo{person}{Tao Yu}, \bibinfo{person}{Toby Jia-Jun Li}, \bibinfo{person}{Shiqi Jiang}, \bibinfo{person}{Yunhao Liu}, \bibinfo{person}{Yaqin Zhang}, {and} \bibinfo{person}{Yunxin Liu}.} \bibinfo{year}{2024}\natexlab{}.
\newblock \showarticletitle{AutoDroid: LLM-powered Task Automation in Android}. In \bibinfo{booktitle}{\emph{Proceedings of the 30th Annual International Conference on Mobile Computing and Networking}} (Washington D.C., DC, USA) \emph{(\bibinfo{series}{ACM MobiCom '24})}. \bibinfo{publisher}{Association for Computing Machinery}, \bibinfo{address}{New York, NY, USA}, \bibinfo{pages}{543–557}.
\newblock
\showISBNx{9798400704895}
\urldef\tempurl%
\url{https://doi.org/10.1145/3636534.3649379}
\showDOI{\tempurl}


\bibitem[Wu et~al\mbox{.}(2024)]%
        {wu2024osatlasfoundationactionmodel}
\bibfield{author}{\bibinfo{person}{Zhiyong Wu}, \bibinfo{person}{Zhenyu Wu}, \bibinfo{person}{Fangzhi Xu}, \bibinfo{person}{Yian Wang}, \bibinfo{person}{Qiushi Sun}, \bibinfo{person}{Chengyou Jia}, \bibinfo{person}{Kanzhi Cheng}, \bibinfo{person}{Zichen Ding}, \bibinfo{person}{Liheng Chen}, \bibinfo{person}{Paul~Pu Liang}, {and} \bibinfo{person}{Yu Qiao}.} \bibinfo{year}{2024}\natexlab{}.
\newblock \bibinfo{title}{OS-ATLAS: A Foundation Action Model for Generalist GUI Agents}.
\newblock
\newblock
\showeprint[arxiv]{2410.23218}~[cs.CL]
\urldef\tempurl%
\url{https://arxiv.org/abs/2410.23218}
\showURL{%
\tempurl}


\bibitem[Xing et~al\mbox{.}(2024)]%
        {xing2024understandingweaknesslargelanguage}
\bibfield{author}{\bibinfo{person}{Mingzhe Xing}, \bibinfo{person}{Rongkai Zhang}, \bibinfo{person}{Hui Xue}, \bibinfo{person}{Qi Chen}, \bibinfo{person}{Fan Yang}, {and} \bibinfo{person}{Zhen Xiao}.} \bibinfo{year}{2024}\natexlab{}.
\newblock \bibinfo{title}{Understanding the Weakness of Large Language Model Agents within a Complex Android Environment}.
\newblock
\newblock
\showeprint[arxiv]{2402.06596}~[cs.AI]
\urldef\tempurl%
\url{https://arxiv.org/abs/2402.06596}
\showURL{%
\tempurl}


\bibitem[Yao et~al\mbox{.}(2023)]%
        {yao2023react}
\bibfield{author}{\bibinfo{person}{Shunyu Yao}, \bibinfo{person}{Jeffrey Zhao}, \bibinfo{person}{Dian Yu}, \bibinfo{person}{Nan Du}, \bibinfo{person}{Izhak Shafran}, \bibinfo{person}{Karthik Narasimhan}, {and} \bibinfo{person}{Yuan Cao}.} \bibinfo{year}{2023}\natexlab{}.
\newblock \showarticletitle{{ReAct}: Synergizing Reasoning and Acting in Language Models}. In \bibinfo{booktitle}{\emph{International Conference on Learning Representations (ICLR)}}.
\newblock


\bibitem[Zhang et~al\mbox{.}(2023)]%
        {zhang2023appagentmultimodalagentssmartphone}
\bibfield{author}{\bibinfo{person}{Chi Zhang}, \bibinfo{person}{Zhao Yang}, \bibinfo{person}{Jiaxuan Liu}, \bibinfo{person}{Yucheng Han}, \bibinfo{person}{Xin Chen}, \bibinfo{person}{Zebiao Huang}, \bibinfo{person}{Bin Fu}, {and} \bibinfo{person}{Gang Yu}.} \bibinfo{year}{2023}\natexlab{}.
\newblock \bibinfo{title}{AppAgent: Multimodal Agents as Smartphone Users}.
\newblock
\newblock
\showeprint[arxiv]{2312.13771}~[cs.CV]
\urldef\tempurl%
\url{https://arxiv.org/abs/2312.13771}
\showURL{%
\tempurl}


\bibitem[Zhang et~al\mbox{.}(2024)]%
        {zhang2024mobileenvbuildingqualifiedevaluation}
\bibfield{author}{\bibinfo{person}{Danyang Zhang}, \bibinfo{person}{Zhennan Shen}, \bibinfo{person}{Rui Xie}, \bibinfo{person}{Situo Zhang}, \bibinfo{person}{Tianbao Xie}, \bibinfo{person}{Zihan Zhao}, \bibinfo{person}{Siyuan Chen}, \bibinfo{person}{Lu Chen}, \bibinfo{person}{Hongshen Xu}, \bibinfo{person}{Ruisheng Cao}, {and} \bibinfo{person}{Kai Yu}.} \bibinfo{year}{2024}\natexlab{}.
\newblock \bibinfo{title}{Mobile-Env: Building Qualified Evaluation Benchmarks for LLM-GUI Interaction}.
\newblock
\newblock
\showeprint[arxiv]{2305.08144}~[cs.AI]
\urldef\tempurl%
\url{https://arxiv.org/abs/2305.08144}
\showURL{%
\tempurl}


\bibitem[Zhang et~al\mbox{.}(2025)]%
        {zhang2025qwen3embeddingadvancingtext}
\bibfield{author}{\bibinfo{person}{Yanzhao Zhang}, \bibinfo{person}{Mingxin Li}, \bibinfo{person}{Dingkun Long}, \bibinfo{person}{Xin Zhang}, \bibinfo{person}{Huan Lin}, \bibinfo{person}{Baosong Yang}, \bibinfo{person}{Pengjun Xie}, \bibinfo{person}{An Yang}, \bibinfo{person}{Dayiheng Liu}, \bibinfo{person}{Junyang Lin}, \bibinfo{person}{Fei Huang}, {and} \bibinfo{person}{Jingren Zhou}.} \bibinfo{year}{2025}\natexlab{}.
\newblock \bibinfo{title}{Qwen3 Embedding: Advancing Text Embedding and Reranking Through Foundation Models}.
\newblock
\newblock
\showeprint[arxiv]{2506.05176}~[cs.CL]
\urldef\tempurl%
\url{https://arxiv.org/abs/2506.05176}
\showURL{%
\tempurl}


\bibitem[Zhang and Zhang(2024)]%
        {zhang2024lookscreensmultimodalchainofaction}
\bibfield{author}{\bibinfo{person}{Zhuosheng Zhang} {and} \bibinfo{person}{Aston Zhang}.} \bibinfo{year}{2024}\natexlab{}.
\newblock \bibinfo{title}{You Only Look at Screens: Multimodal Chain-of-Action Agents}.
\newblock
\newblock
\showeprint[arxiv]{2309.11436}~[cs.CL]
\urldef\tempurl%
\url{https://arxiv.org/abs/2309.11436}
\showURL{%
\tempurl}


\bibitem[Zheng et~al\mbox{.}(2024)]%
        {zheng2024seeact}
\bibfield{author}{\bibinfo{person}{Boyuan Zheng}, \bibinfo{person}{Boyu Gou}, \bibinfo{person}{Jihyung Kil}, \bibinfo{person}{Huan Sun}, {and} \bibinfo{person}{Yu Su}.} \bibinfo{year}{2024}\natexlab{}.
\newblock \showarticletitle{GPT-4V(ision) is a Generalist Web Agent, if Grounded}. In \bibinfo{booktitle}{\emph{Forty-first International Conference on Machine Learning}}.
\newblock
\urldef\tempurl%
\url{https://openreview.net/forum?id=piecKJ2DlB}
\showURL{%
\tempurl}


\end{thebibliography}

\end{document}